\documentclass[12pt,preprint]{aastex}
\newcommand{\beq}{\begin{equation}}
\newcommand{\beqa}{\begin{eqnarray}}
\newcommand{\eeq}{\end{equation}}
\newcommand{\eeqa}{\end{eqnarray}}

\def\f{\frac}

\def\hess{{{HESS~J1745$-$303~}}}

\slugcomment{submitted to ApJ}
\shorttitle{HESS~J1745$-$303 with {\it Suzaku}}
\shortauthors{Bamba et al.}
\begin{document}


\title{X-ray Observation of Very High Energy Gamma-ray Source,
HESS~J1745$-$303, with {\it Suzaku}}

\author{
Aya~Bamba\altaffilmark{1},
Ryo~Yamazaki\altaffilmark{2}, 
Kazunori~Kohri\altaffilmark{3},
Hironori~Matsumoto\altaffilmark{4},
Stefan~Wagner\altaffilmark{5},
Gerd P\"{u}hlhofer\altaffilmark{5},
Karl Kosack\altaffilmark{6}
}

\altaffiltext{1}{
ISAS/JAXA Department of High Energy Astrophysics
3-1-1 Yoshinodai, Sagamihara,
Kanagawa 229-8510, JAPAN;
bamba@astro.isas.jaxa.jp
}

\altaffiltext{2}{
Department of Physical Science, Hiroshima University, Higashi-Hiroshima,
Hiroshima 739-8526
}

\altaffiltext{3}{
Physics Department, Lancaster University, Lancaster, LA1~4YB, UK
}

\altaffiltext{4}{
Department of Physics, Kyoto University, Kyoto 606-8502, Japan
}

\altaffiltext{5}{
Landessternwarte, Heidelberg, K\"{o}nigstuhl 12, 69117, Heidelberg, Germany
}

\altaffiltext{5}{
Max-Planck-Institut f\"{u}r Kernphysik, Heidelberg, Germany}

\begin{abstract}
{\it Suzaku} observations of a TeV unidentified (unID) source,
HESS~J1745$-$303, are presented.
A possible excess of neutral iron line emission is discovered, and
 is likely associated with 
the main part of HESS~J1745$-$303, named ``region~A''.
It may be an X-ray reflection nebula where the X-rays from
previous Galactic Center (GC) activity are reflected by
a molecular cloud.
This result further strengthens
the assumption that the molecular cloud which is 
spatially coincident with region A of HESS~J1745$-$303 is located 
in the GC region.
The TeV emission from molecular clouds is reminiscent of 
the diffuse TeV gamma-rays from the GC giant molecular clouds,
and it could have the same emission mechanism.
With deep exposure mapping observations by {\it Suzaku},
a tight upper-limit on the 2--10~keV continuum
diffuse emission from region A is obtained,
as $2.1\times 10^{-13}$~ergs~s$^{-1}$cm$^{-2}$.
The flux ratio between 1--10~TeV and 2--10~keV is larger than 4.
Possible scenarios to reproduce wide-band spectra from keV to TeV
are examined.
Thermal X-rays from nearby two old supernova remnants,
G359.0$-$0.9 and G359.1$-$0.5, are detected,
and their emission properties
are well determined in the present study
with deep exposure.
\end{abstract}



\section{Introduction}
\label{sec:intro}

Very high-energy gamma-rays with energies around TeV
are powerful tools to investigate the cosmic accelerators.
They arise from either leptonic
(cosmic microwave background (CMB) or other soft seed photons 
up-scattered by accelerated electrons and their bremsstrahlung emission)
or hadronic (the
decay of neutral pions, arising from the collision of high energy
protons and interstellar matter) processes.
The strong evidence for electron acceleration up to more than 
$\sim10$~TeV has been obtained with X-ray observations 
of, for example, young supernova remnants (SNRs) that are
believed to be the most probable cosmic-ray accelerators in our Galaxy
\citep[e.g.,][]{koyama1995,koyama1997,
bamba2003a,bamba2003b,bamba2005a,bamba2005b}.
On the other hand, we have not yet derived firm evidence for 
proton acceleration; although TeV gamma-rays have
been detected by several of young SNRs, such as 
RX~J1713.7$-$3946 \citep{enomoto2002,aharonian2004}
and RX~J0852.0$-$4622 \citep{katagiri2005,aharonian2005c},
their origin is not yet fully understood
\citep[e.g,][]{pannuti2003,lazendic2004,ellison2001,bamba2005b,
uchiyama2005,uchiyama2007,yamazaki2008}.

Recently, a survey of the inner part of our Galaxy
has revealed several new TeV $\gamma$-ray sources 
\citep{aharonian2002,aharonian2005,aharonian2005b,aharonian2006}. 
Some of them have no counterpart in the X-ray regime.
Then, the flux ratio, defined by
\[
R_{\rm TeV/X}=\f{F_\gamma(1-10~{\rm TeV})}{F_X(2-10~{\rm keV})}~~,
\]
is more than $\sim10$ \citep{yamazaki2006,bamba2007a,matsumoto2007}.
On the other hand, young supernova remnants (SNRs) and pulsar wind nebulae 
have $R_{\rm TeV/X}$ less than $\sim2$.
It is interesting that
the TeV unidentified sources with large $R_{\rm TeV/X}$
may show evidence for
hadron acceleration because the leptonic inverse-Compton model requires
an unusually small magnetic field strength ($\ll1~\mu$G).
So far, several TeV gamma-ray objects have been followed up
in X-rays, which provide us a lot of hint
on the emission process of TeV gamma-rays 
\citep[e.g.,][]{tian2007,bamba2007a,matsumoto2007,matsumoto2008}.

\hess is one of such TeV unidentified sources.
It was discovered by the Galactic plane survey of 
High Energy Stereoscopic System (H.E.S.S.)
\citep{aharonian2006}, and subsequent follow-up observation
was performed recently \citep{aharonian2008}.
The observed flux from the whole object is estimated as 
$F_\gamma(1-10~{\rm TeV})\sim5.2\times10^{-12}$~erg~s$^{-1}$cm$^{-2}$
The low Galactic latitude and the fact that the source is spatially
extended strongly argues for a Galactic origin of \hess.
\hess  mainly consists of three bright emission regions
\citep[named region~A, B, and C;][]{aharonian2008},
and the angular size of the whole object is  
$\sim0.3^\circ\times0.5^\circ$,
which is one of the largest TeV unID sources.
\hess has direct positional coincidence with an unidentified EGRET
source \citep[3EG~J1744$-$3011,][]{hartman1999}, which implies
proton acceleration because the leptonic IC model for the TeV
emission usually predict very dim GeV $\gamma$-rays.
\citet{aharonian2008} suggested that
a molecular cloud coincides with the TeV emitting region,
although the precise distance to the molecular cloud is unknown.
This region was also observed with {\it XMM-Newton},
which gives an upper-limit on the diffuse X-ray emission,
$4.5\times10^{-13}$~erg~cm$^{-2}$s$^{-1}$,
 from the region~A which is the main part of \hess and
has a TeV energy flux of 
$8.0\times10^{-13}$~erg~cm$^{-2}$s$^{-1}$.
Then, the TeV-to-X-ray flux ratio, $R_{\rm TeV/X}$, is larger 
than $\sim2$.

In this paper, we report on the deep mapping observations of \hess
with X-ray satellite {\it Suzaku} \citep{mitsuda2007}.
Section~\ref{sec:obs} introduces the observations. Our analysis
is explained in \S~\ref{sec:analysis}, and we
discuss our findings in \S~\ref{sec:discuss}.

\section{Observations and Data Reduction}
\label{sec:obs}

We carried out mosaic observations of HESS~J1745$-$303
with {\it Suzaku}.
The mapping used four pointings as listed in Table~\ref{tab:obslog}.
{\it Suzaku} has two active instruments,
four X-ray Imaging Spectrometers \citep[XIS0--XIS3;][]{koyama2007}
each at the focus of an X-Ray Telescope \citep[XRTs][]{serlemitsos2007}
and a separate Hard X-ray Detector \citep[HXD;][]{takahashi2007}.
All four spectrometers were active in the first observation,
whereas only three could be used subsequently
due to a problem with XIS~2.
XIS1 is a back-illuminated (BI) CCD,
whereas the others are front-illuminated (FI).
The XIS was operated
in the normal full-frame clocking mode.
The spaced-row charge injection \citep{nakajima2008}
was used in the later three observations (Table~\ref{tab:obslog}).
The data reduction and analysis were made
using HEADAS software version 6.4.,
version 2.2.7.18 of the processed data,
and XSPEC version 11.3.2.
We filtered out data obtained during passages through the
South Atlantic Anomaly (SAA),
with elevation angle to the Earth's limb below 5$^\circ$,
or with elevation angle to the bright Earth's limb below 13$^\circ$.
in order to avoid the contamination of emission from the bright Earth.
In this paper, we concentrate on the XIS data analysis,
because HESS~J1745$-$303 is located near the Galactic plane,
and the HXD data is 
contaminated by the Galactic ridge X-ray emission
\citep[GRXE; c.f.,][]{koyama1989,yuasa2007}
and bright point sources,
which are quite difficult to remove correctly.

\section{Results}
\label{sec:analysis}

\subsection{Images}

We made mosaic intensity maps in the 0.5--2.0~keV,
2.0--8.0~keV, and 6.3--6.5~keV bands
of the four pointing observations as described in
the following.

The raw mosaic images were made in each energy band.
In order to estimate the continuum emission in the narrow iron intensity band,
we derived the spectrum over the entire region
and fitted 3--8~keV photons
with a power-law plus narrow Gaussians
which represent emission lines.
The 2--3~keV band photons were not used in order to avoid the influence of
soft discrete emission (two SNRs; described later).
The best-fit photon index was $\sim 1--1.3$.
We compared the count rates of the power-law component
in the 2.0--8.0~keV and 6.3--6.5~keV bands,
the latter is 1.15\% of the former,
and normalized the 2.0--8.0~keV image with the ratio.
The normalized continuum image was subtracted from the
6.3--6.5~keV image, which is the ``neutral iron intensity image''.

The non-X-ray background (NXB) image
for each energy band was generated from the NXB database
provided by the Suzaku team.
The NXB images were chosen to make the average geomagnetic cu-off rigidity
same as that of the observation,
and extracted from the same region in the detector.
In this procedure,
{\tt xisnxbgen} package developed by \citet{tawa2008} was used.
After subtracting NXB images from the raw images,
we combined the NXB-subtracted images using {\tt ximage} package.

\begin{figure}
\epsscale{0.45}
\plotone{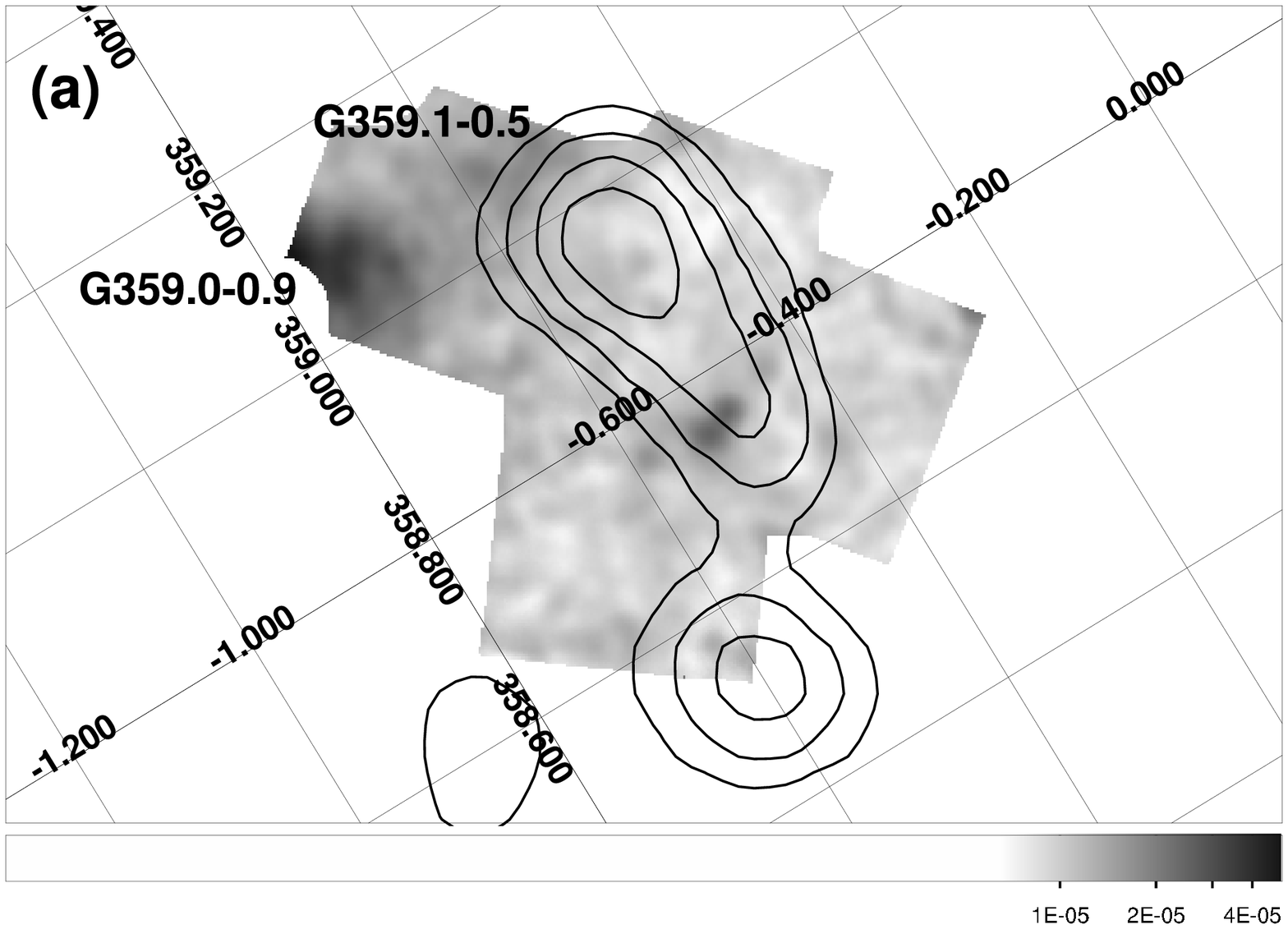}
\plotone{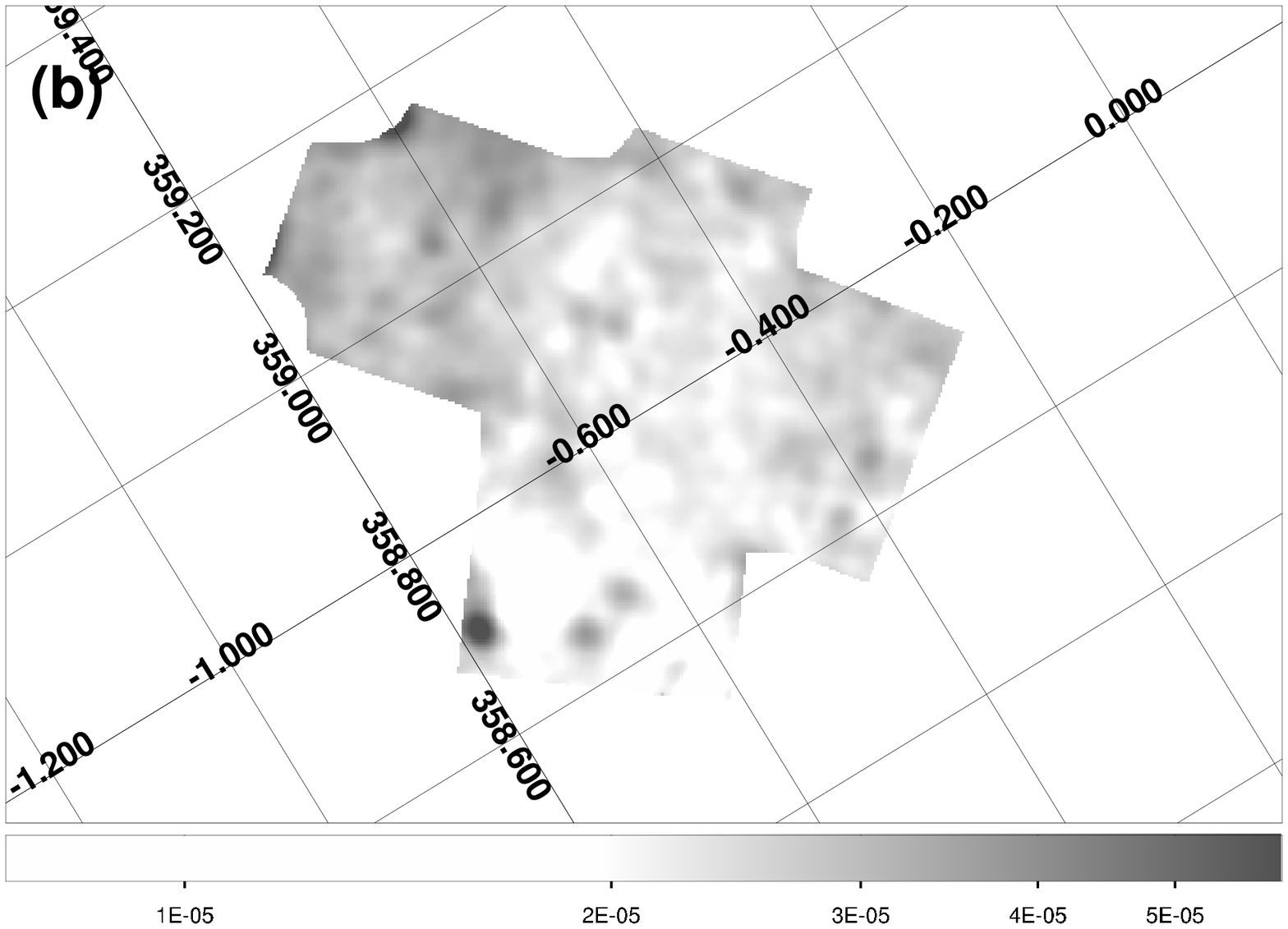}
\plotone{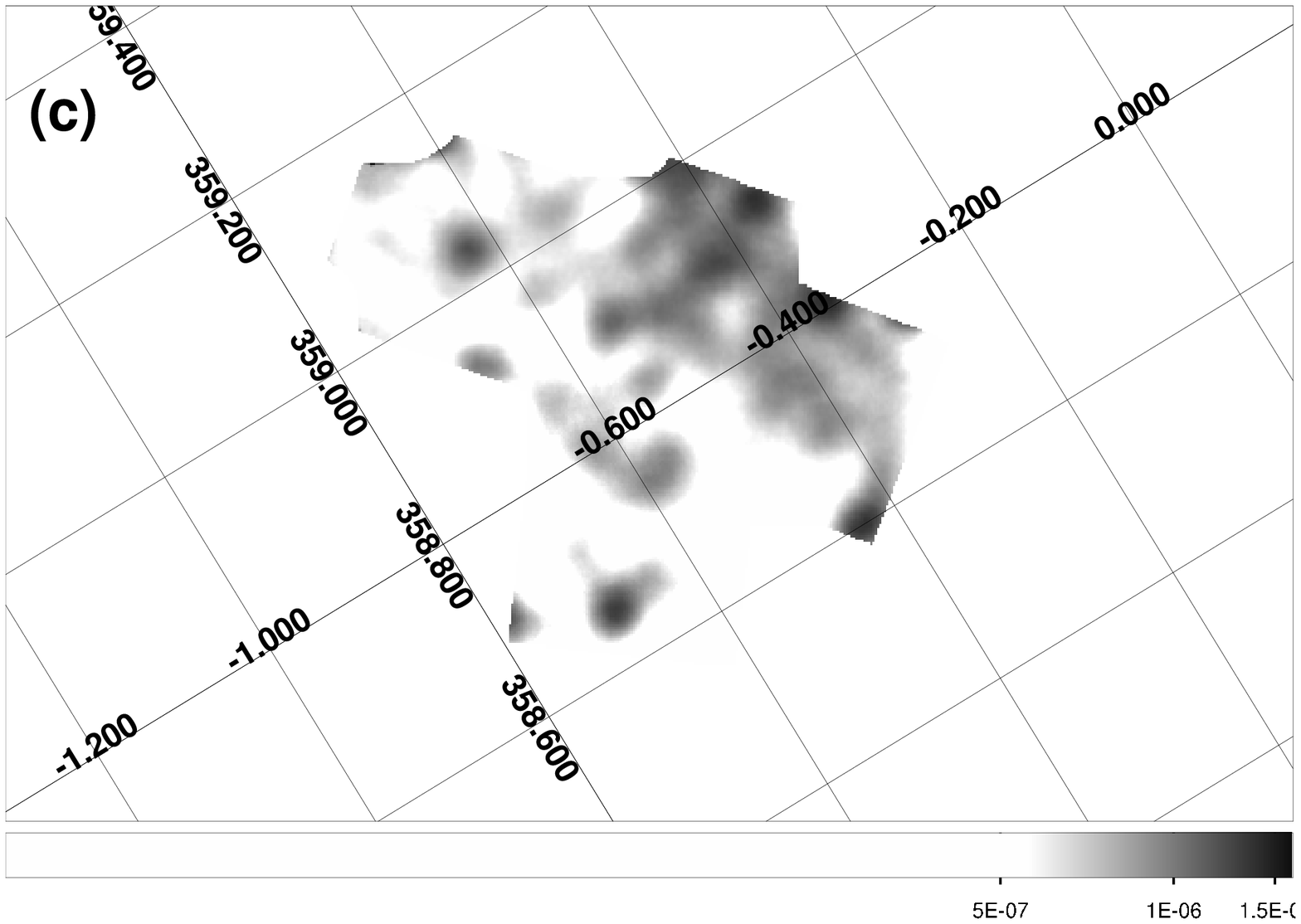}
\caption{{\it Suzaku} XIS intensity maps of the HESS~J1745$-$303 region
in the 0.5--2.0~keV (a) and 2.0--8.0~keV (b) bands,
and the neutral iron line (c).
The scale is in the logarithmic.
The contour represents TeV emission in the linear scale.}
\label{fig:images}
\end{figure}

The exposure maps were made 
for 0.5--2.0 and 2.0--8.0~keV bands
with the {\tt xissim} package \citep{ishisaki2007}.
For the neutral iron line map,
the 2.0--8.0~keV energy range was used.
The exposure maps were also combined.
Dividing the NXB-subtracted images with the exposure maps,
we made the mosaic images
shown in Figure~\ref{fig:images}.
Our observations cover  all of the region~A of \hess.

On the eastern edge of \hess,
we can see some diffuse components in the soft X-ray band,
(see Figure~\ref{fig:images}(a)).
They coincide with two radio SNRs,
G359.0$-$0.9 and G359.1$-$0.5
\citep{larosa2000,bamba2000}.
Some point sources are also seen, 
as reported by \citet{aharonian2008}.
However, no excess was found within the \hess region.
The hard band image (Figure~\ref{fig:images}b) also shows no diffuse source.
In the neutral iron line map (Figure~\ref{fig:images}c), on the other hand,
some excess is shown around region~A.

\subsection{Spectra}

\subsubsection{The X-ray spectrum of the region~A}

\begin{figure}
\epsscale{0.45}
\plotone{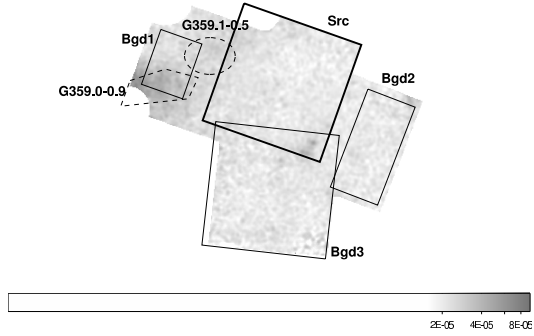}
\caption{Regions for the spectral analysis overlaid on the
0.5--2.0~keV image.
Thick and thin solid rectangles are source and background regions
for the HESS~J1745$-$303 region analysis.
Dashed regions are for G359.0$-$0.9 and G359.1-0.5 analysis.}
\label{fig:regions}
\end{figure}

We selected source and background regions (``Src'' and ``Bgd1--3'' hereafter) 
as shown with solid lines in Figure~\ref{fig:regions},
to make the upper-limit of the X-ray emission
of HESS~J1745$-$303 region~A.
The background regions were divided to 3 regions
(Bgd1--3 in Figure~\ref{fig:regions}).
Two types of background should be considered,
NXB and the GRXE
\citep[for example]{yamauchi1993}.
The latter has positional dependence along the Galactic longitude.
We thus used Bgd1 and 2 regions
to make the average longitude of each region similar.
The 3--8~keV count rates of XIS~0 
are $(6.0\pm 0.05)\times 10^{-4}$
and $(7.0\pm 0.1)\times 10^{-4}$~cnts~s$^{-1}$arcmin$^{-2}$
for the source and  the background regions, respectively.
For making the upper-limit of the X-ray flux
in the source region,
the response and auxiliary files were constructed using
{\tt xisrmfgen} and {\tt xissimarfgen} \citep{ishisaki2007}
in the HEADAS package,
with the assumption that
the emission comes from the source region uniformly.
The photons from the background regions were subtracted
from the source spectra.
We fit the background-subtracted spectra
in the 2--10~keV band
with a power-law model with fixed $\Gamma = 2$,
which is significantly harder value than
the TeV emission \citep{aharonian2008},
and to derive the 90\% upper-limit of
2.1$\times 10^{-13}$~ergs~cm$^{-2}$s$^{-1}$.
The larger photon indices were also used to estimate
the systematic errors of the upper-limit
and derived the smaller value.

\begin{figure}
\epsscale{0.45}
\plotone{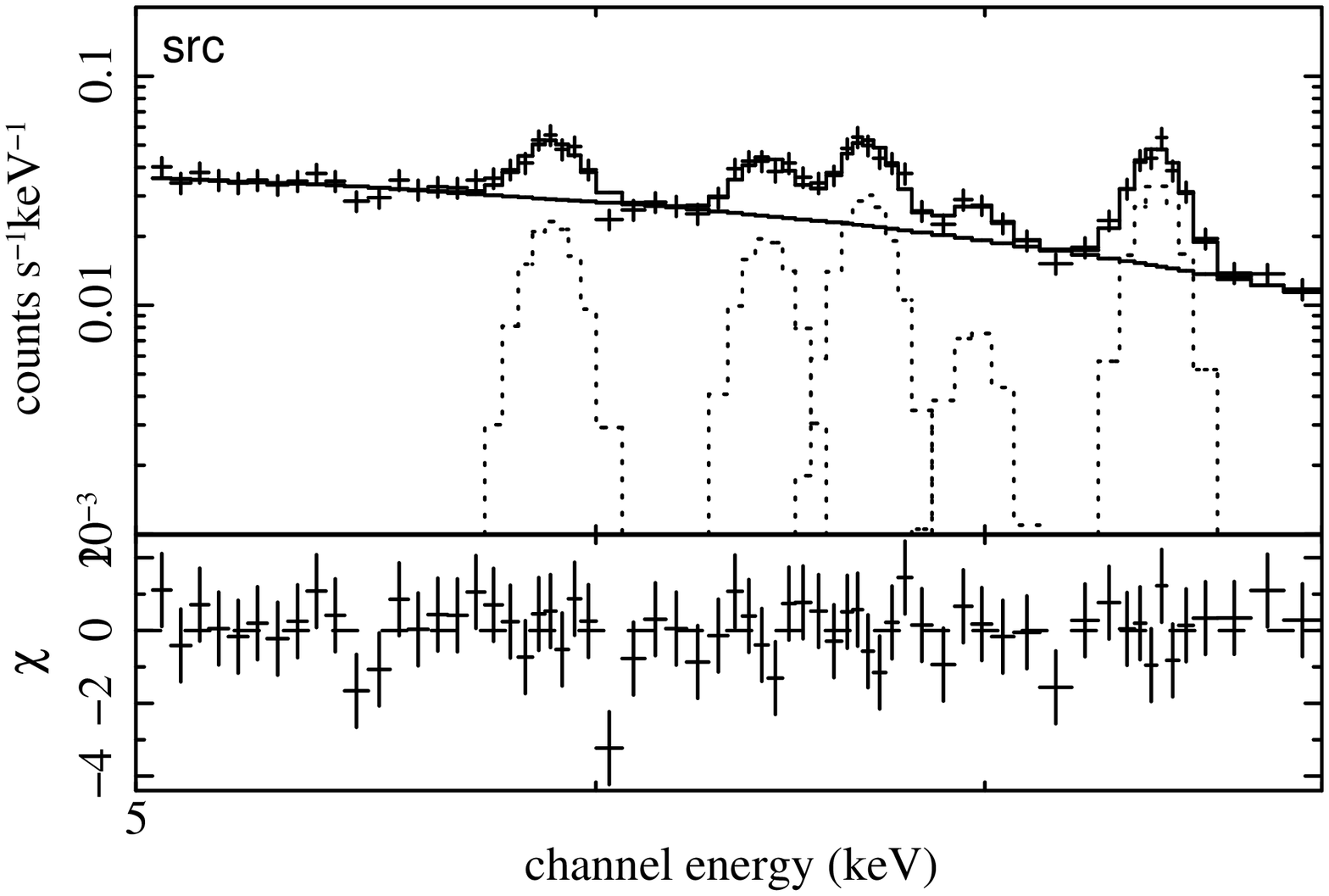}
\plotone{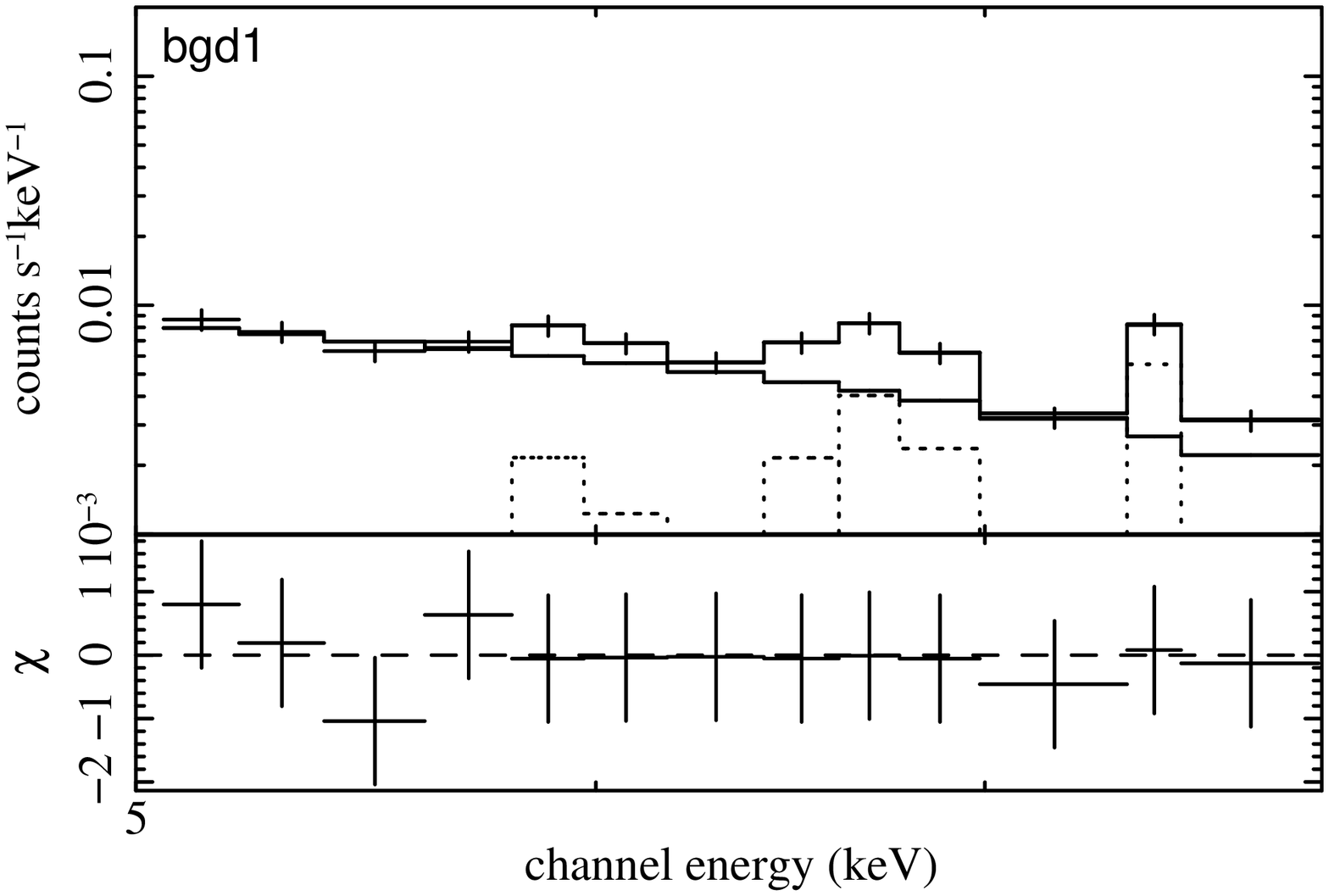}
\plotone{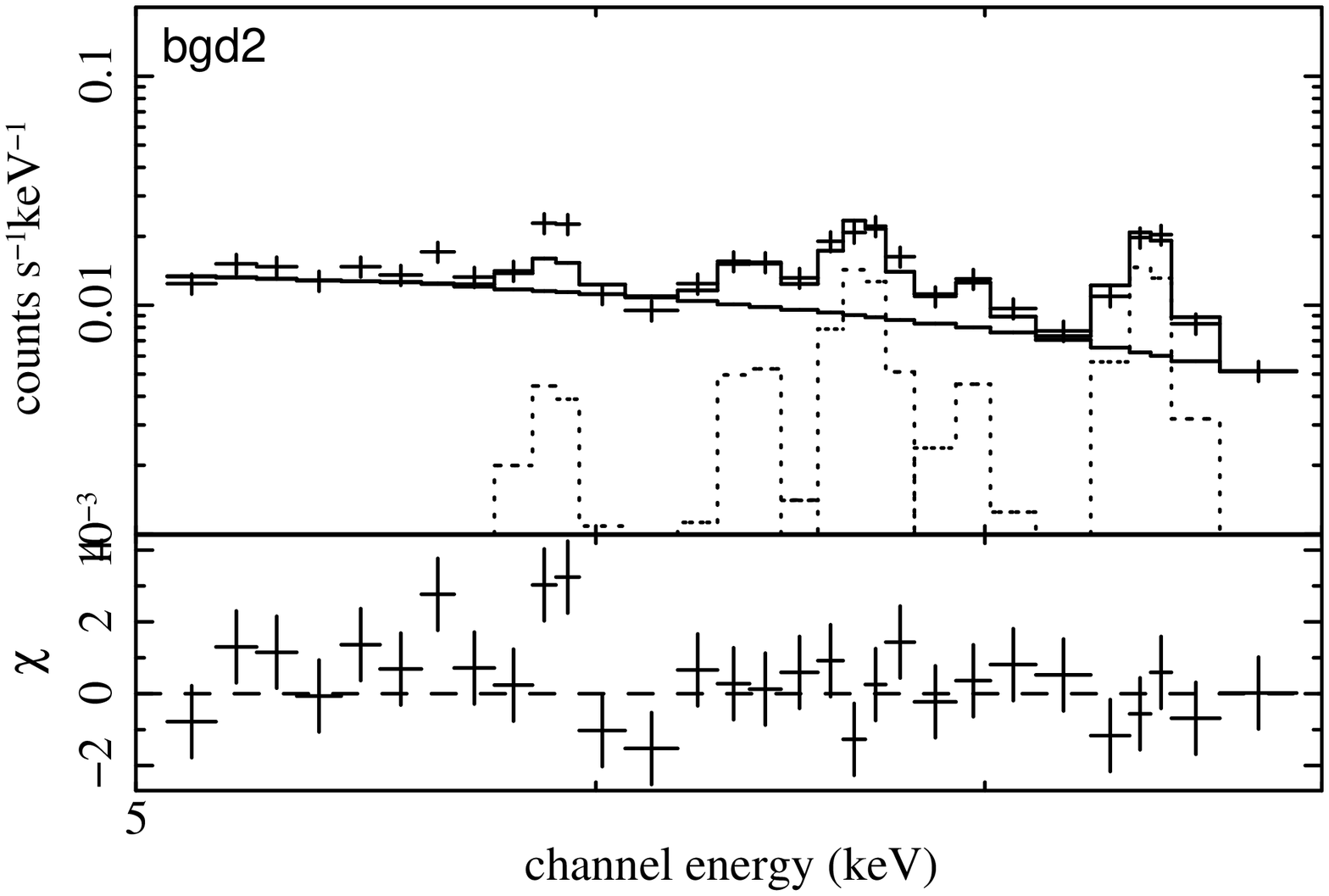}
\plotone{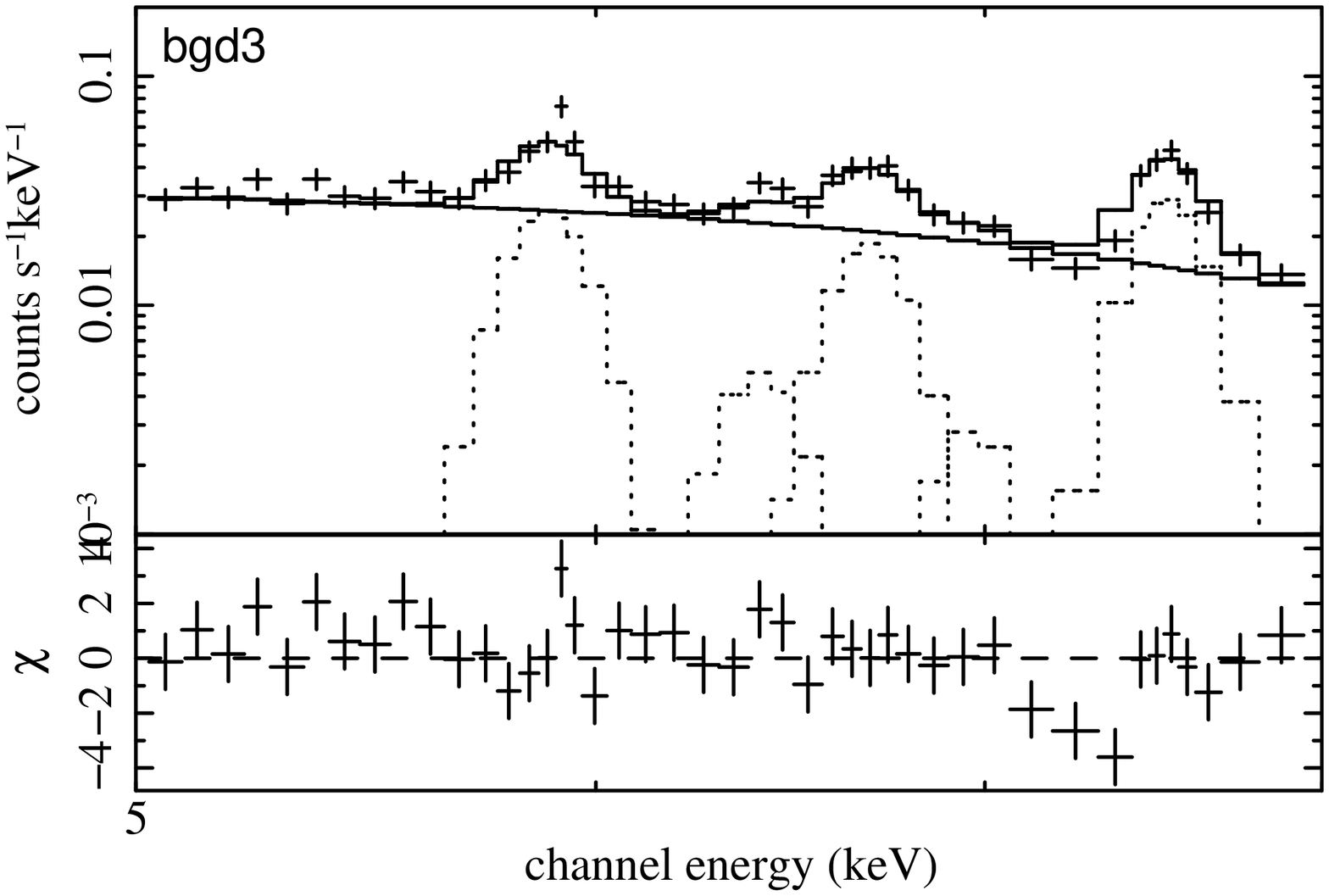}
\caption{XIS spectra of src and bgd regions.
Background photons were not subtracted.
Solid and dotted lines represents
the continuum and emission lines.
Although the fittings were done with 4 XISs,
only XIS0 data are plotted for convenience.
}
\label{fig:spectra}
\end{figure}

We also derived the surface brightness of the neutral iron emission line
in each region.
Only the FI CCD data were used since the BI CCD has larger background
in the hard X-ray band.
We fitted 5--8~keV spectra with a power-law component, 
narrow Gaussians for neutral, He-like, and H-like iron lines,
and for Mn-K, and Ni-K lines as NXB emission.
The best-fit models and parameters of the neutral and He-like iron lines 
in each region are
shown in Figure~\ref{fig:spectra} and Table~\ref{tab:iron}.
The surface brightness of He-like iron is almost constant
between these regions.
We added all background regions to increase statistics
and derived the average values of line intensity.
After the procedure,
the neutral iron line is significantly stronger in the source region
than those in background regions,
whereas the He-like iron line is compatible in both regions.
The total excess of the line in these regions
is $1.1\times 10^{-5}$~ph~s$^{-1}$cm$^{-2}$.

\subsubsection{G359.0$-$0.9 and G359.1$-$0.5}

\begin{figure}
\epsscale{0.45}
\plotone{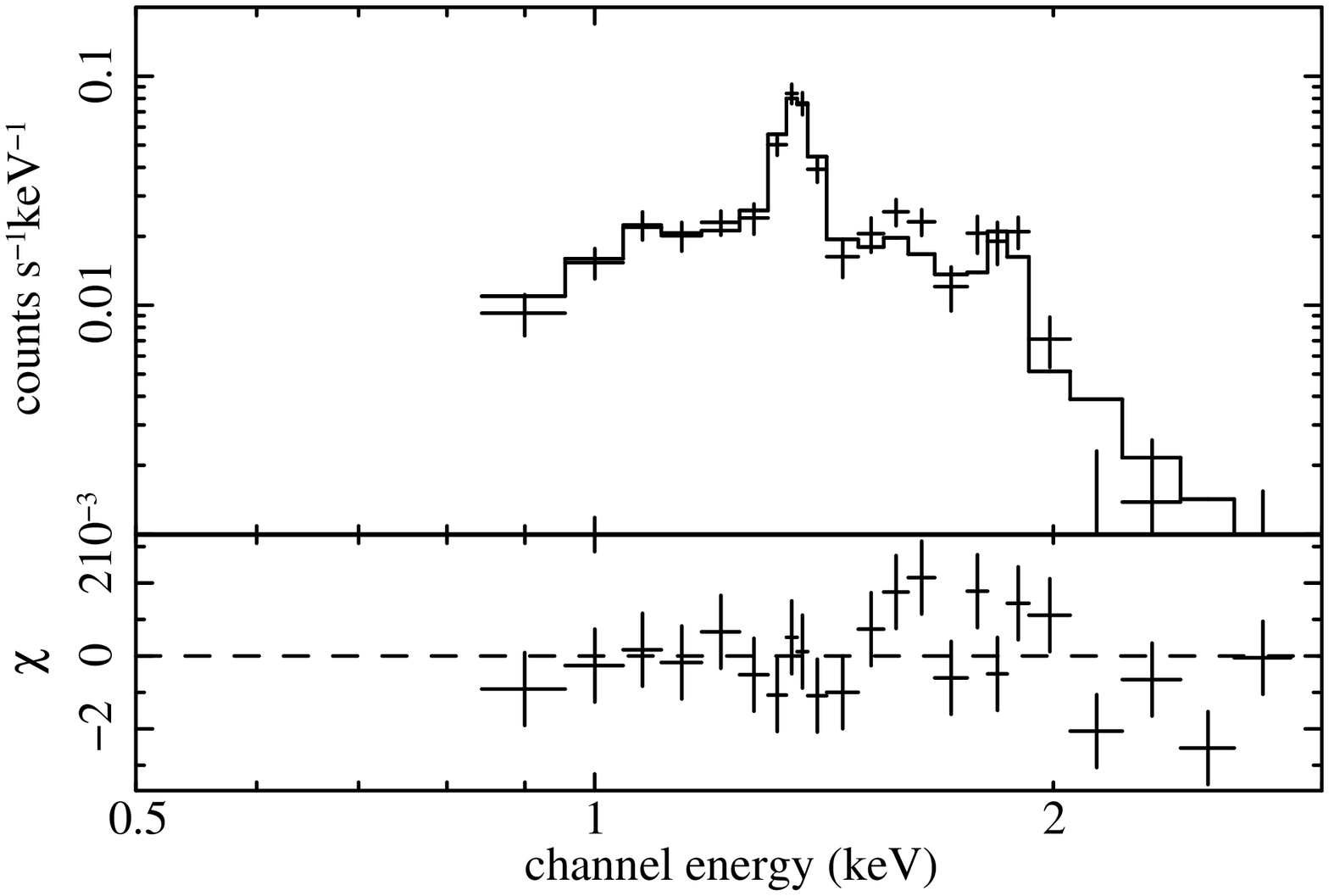}
\plotone{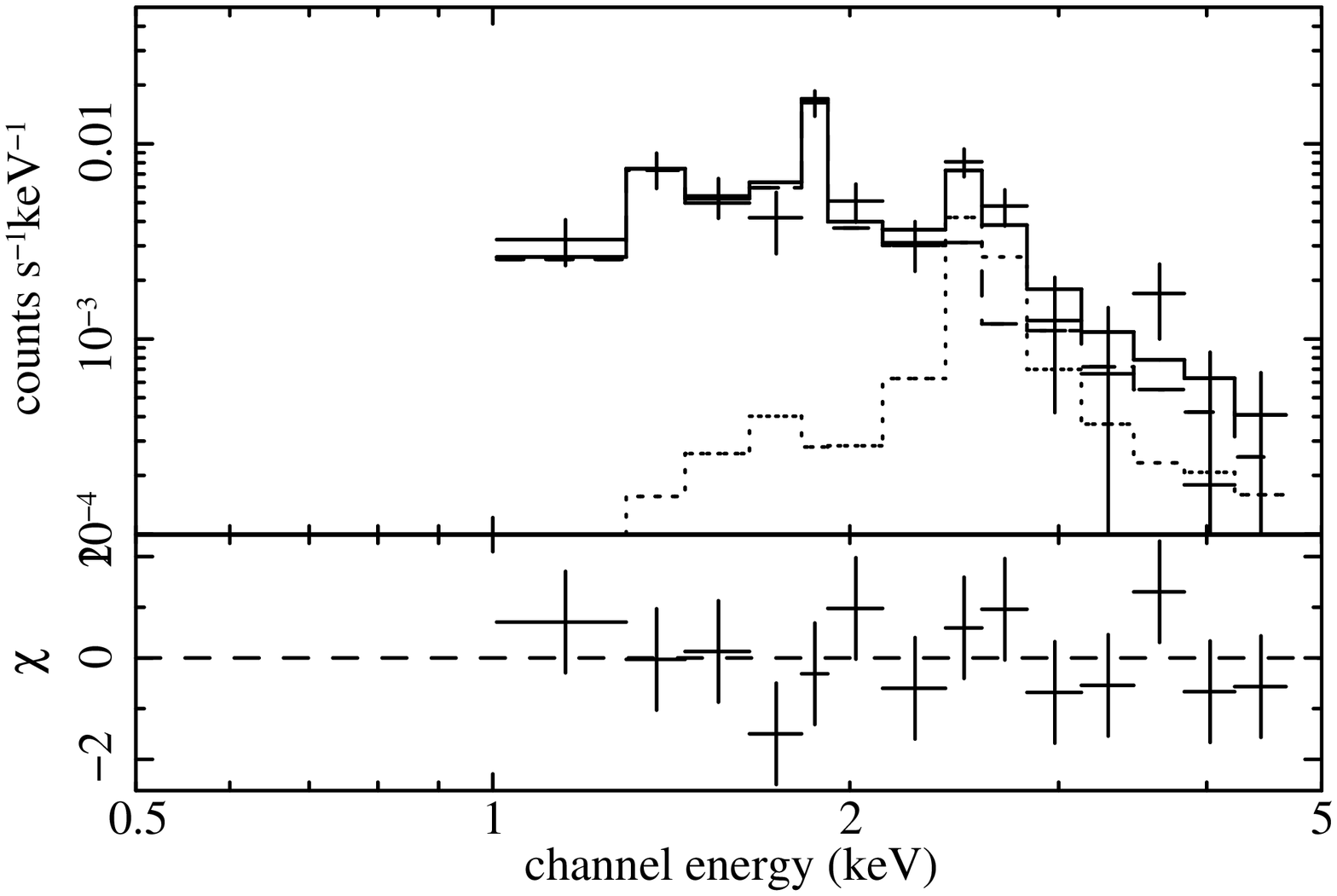}
\caption{XIS Spectra of G359.0$-$0.9 (left) and G359.1$-$0.5 (right).
Only XIS0 data are plotted for convenience.
Solid and dashed lines represents
the lower and higher temperature plasma models.}
\label{fig:snrs}
\end{figure}

For the spectral analysis of the two SNRs,
we used the regions with dashed-lines in Figure~\ref{fig:regions}.
Background photons were accumulated from the source-free regions
in the same observation.
The background-subtracted spectra are shown in Figure~\ref{fig:snrs}.
The spectra are rather soft
and there are emission lines from highly ionized Si and S
in both spectra.
These facts indicate that these emissions are thermal,
which is consistent with the ASCA results \citep{bamba2000}.
We then fitted these spectra
with non-equilibrium ionization collisional plasma model
\citep[{\tt nei} in XSPEC;][]{borkowski2001}
with absorption.
For the absorption model,
we used the cross sections of
\citet{morrison1983} and assumed solar abundances \citep{anders1989}.
The XIS team reported the gain uncertainty of a few eV
\citep[c.f.,][]{koyama2007},
thus the gain is allowed to vary within a few percent separately.
\citet{bamba2000} reported that
the spectrum of G359.1$-$0.5 was well described with
the two-temperature model
with high-S abundance in the higher temperature component.
So we have added another thermal component in the present analysis.
We treated some abundances in Table~\ref{tab:snrs} as free parameters 
and fixed the others at 1 solar.
The best-fit models and parameters are shown in Figure~\ref{fig:snrs}
and Table~\ref{tab:snrs}.

The results are consistent with the {\it ASCA} results,
but the abundance and the ionization time scale are better determined
thanks to the better energy resolution of XIS.
We expect a difference in flux,
since our observations do not cover the entire remnants.
The absorption column of G359.0$-$0.9 is smaller than
sources in the Galactic center (GC), thus
this SNR is likely a foreground source.
On the other hand, G359.1$-$0.5 shows a very large
absorption column, which implies
that it is in the GC region.

\section{Discussion}
\label{sec:discuss}

\subsection{The neutral iron line and molecular clouds}

\citet{aharonian2008} reported that HESS~J1745$-$303 region~A
had spatial coincidence with a molecular cloud \citep{bitran1997}.
We suggest that the TeV emission is also associated with
the neutral iron line emission,
although we cannot unambiguously conclude that due to the poor statistics.
The positional coincidence of the neutral iron line and the molecular cloud
reminds us of reflected
X-rays originating in the past activity in the GC
\citep[e.g.,][]{inui2008}.
If the molecular cloud is optically thin,
the line emission intensity, $I_{\rm Fe}$,
should be proportional to the mass of the molecular cloud
and the viewing angle from the GC, so that
\[
I_{\rm Fe} \propto 
\frac{M_{\rm MC}}{(D\theta)^2} \ \ ,
\]
where $M_{\rm MC}$, $\theta$, and $D$ are the mass of the molecular cloud,
the angular size of the molecular cloud viewed from the GC,
and the angular separation between the cloud and the GC,
respectively \citep{nobukawa2008}.
We compared the masses of the molecular cloud
coincident with source~A of HESS~J1745$-$303
and the most famous X-ray reflection nebula,
Sgr~B2 \citep{murakami2001}.
The latter shows neutral iron line emission with an intensity of
$5.6\times 10^{-5}$~ph~s$^{-1}$cm$^{-2}$.
Table~\ref{tab:MC} shows the characteristics of these molecular clouds.
For the molecular cloud at region~A of HESS~J1745$-$303, we obtain
$I_{\rm Fe}\sim7\times10^{-6}$~ph~s$^{-1}$cm$^{-2}$
on the assumption of time-independent ejected X-ray luminosity at the GC.
This estimated value is compatible with the observed one,
$1.1\times 10^{-5}$~ph~s$^{-1}$cm$^{-2}$.
Therefore, we conclude that this molecular cloud is a new example of
X-ray reflection nebula 
and that the molecular cloud is located in the GC region.
We believe that hard X-ray observations are the strong tools
to judge whether the molecular clouds are in the GC region or not. 

\subsection{The upper-limit of the nonthermal Component of X-rays and
               Possible Emission Mechanism of TeV Gamma-rays}

We derive a 90\% X-ray flux upper-limit for the HESS~J1745$-$303 region~A
in the 2--10~keV band of $2.1\times 10^{-13}$~ergs~s$^{-1}$cm$^{-2}$,
which is tighter than the {\it XMM} results \citep{aharonian2008}.
The flux ratio between 2--10~keV and 1--10~TeV,
$R_{\rm TeV/X}$, is larger than 4,
which could be larger than
TeV emitting shell-type SNRs and PWNe
\citep{bamba2007a,matsumoto2007,yamazaki2006}.

In the following,
 we discuss  specific models of the emission mechanism 
to reproduce the observed results of X-ray and TeV gamma-rays for
the region A of HESS~J1745$-$303.
In particular, we focus on whether or not
the upper limit of nonthermal X-ray emission,
 which is derived in the present work, 
gives us significant constraints on the theoretical models.
We consider hadronic and leptonic models separately.
Since \hess is near  the Galactic center, 
the GeV emission may be  overlaid with the intense Galactic diffuse
emission and uncertainty is very large. Hence we neglect it here
for simplicity, and take into account 
only  X-ray and TeV gamma-ray bands.

\subsubsection{Hadronic gamma-ray emission model}
\label{sec:hadronic}

In hadronic models,
TeV gamma-rays originate from accelerated protons.
It is assumed that the proton energy spectrum has a form of
$E^{-p}\exp(-E/E_{\rm max,p})$ with an index, $p$, and
cut-off energy, $E_{\rm max, p}$.
Here, the value of $p$ is highly uncertain,
so we adopt $p=2$ as a typical value.
The high-energy protons collide with target protons in the
ISM or molecular cloud, generating neutral ($\pi^0$)
and charged ($\pi^{\pm}$) pions.
The former decays into gamma rays, while the decay
of the latter causes secondary electrons and positrons
producing bremsstrahlung, inverse Compton,
 and synchrotron emission.

Although it may be likely that the region~A
of \hess is associated with a 
molecular cloud and target matter density
$n$ is high, the low-$n$ case cannot be, at present, excluded. 
So, we consider two cases, 
in which $n=1$~cm$^{-3}$ and $n=5\times10^3$~cm$^{-3}$,
independently.
The value of $n$ in the latter case is adopted according to
\citet{aharonian2008}.
Generally speaking, in the former, bremsstrahlung emission
is not dominant in the TeV gamma-ray band, while it may often be dominant
in the latter.

\begin{figure}
\epsscale{0.45}
\plotone{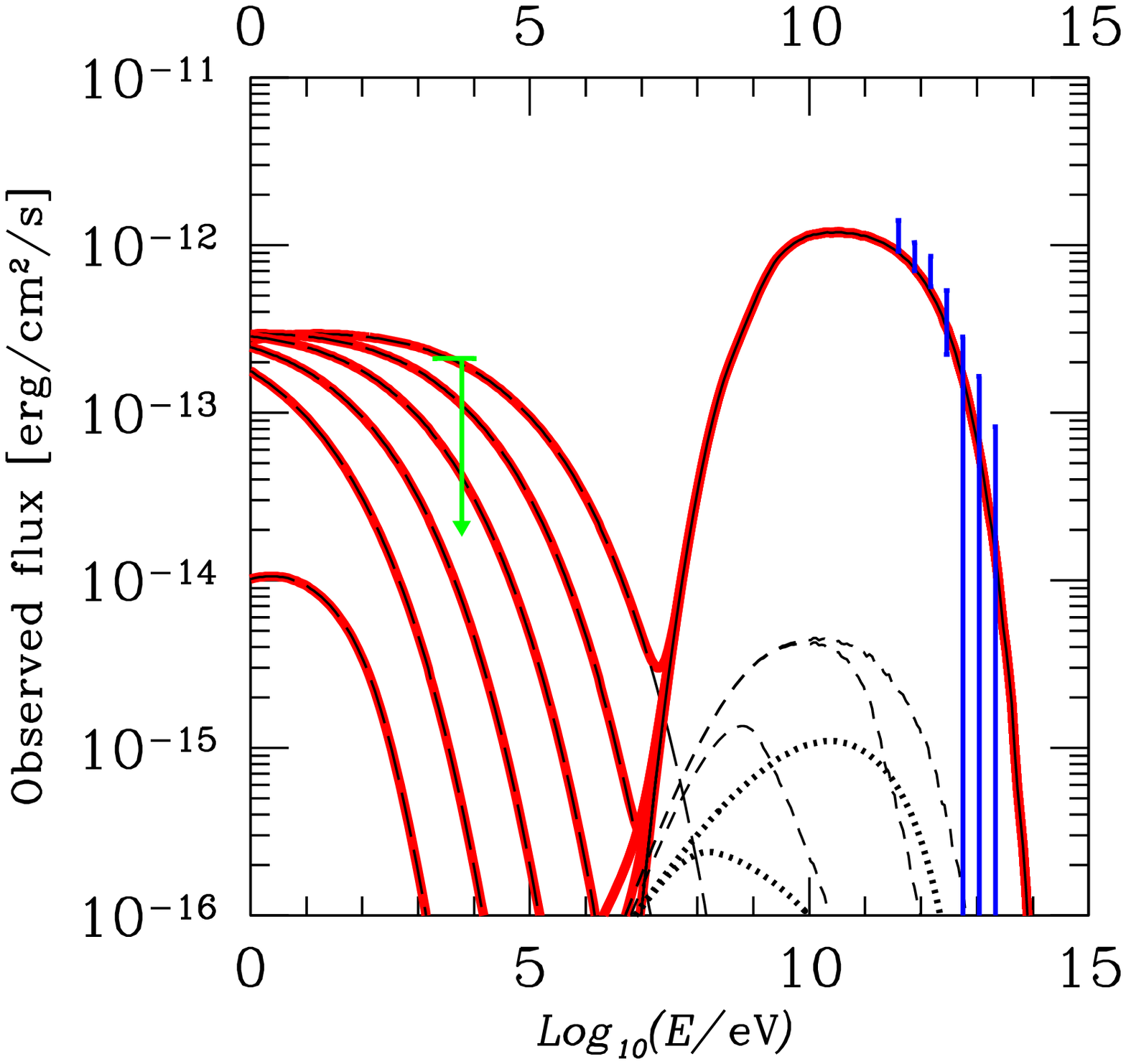}
\plotone{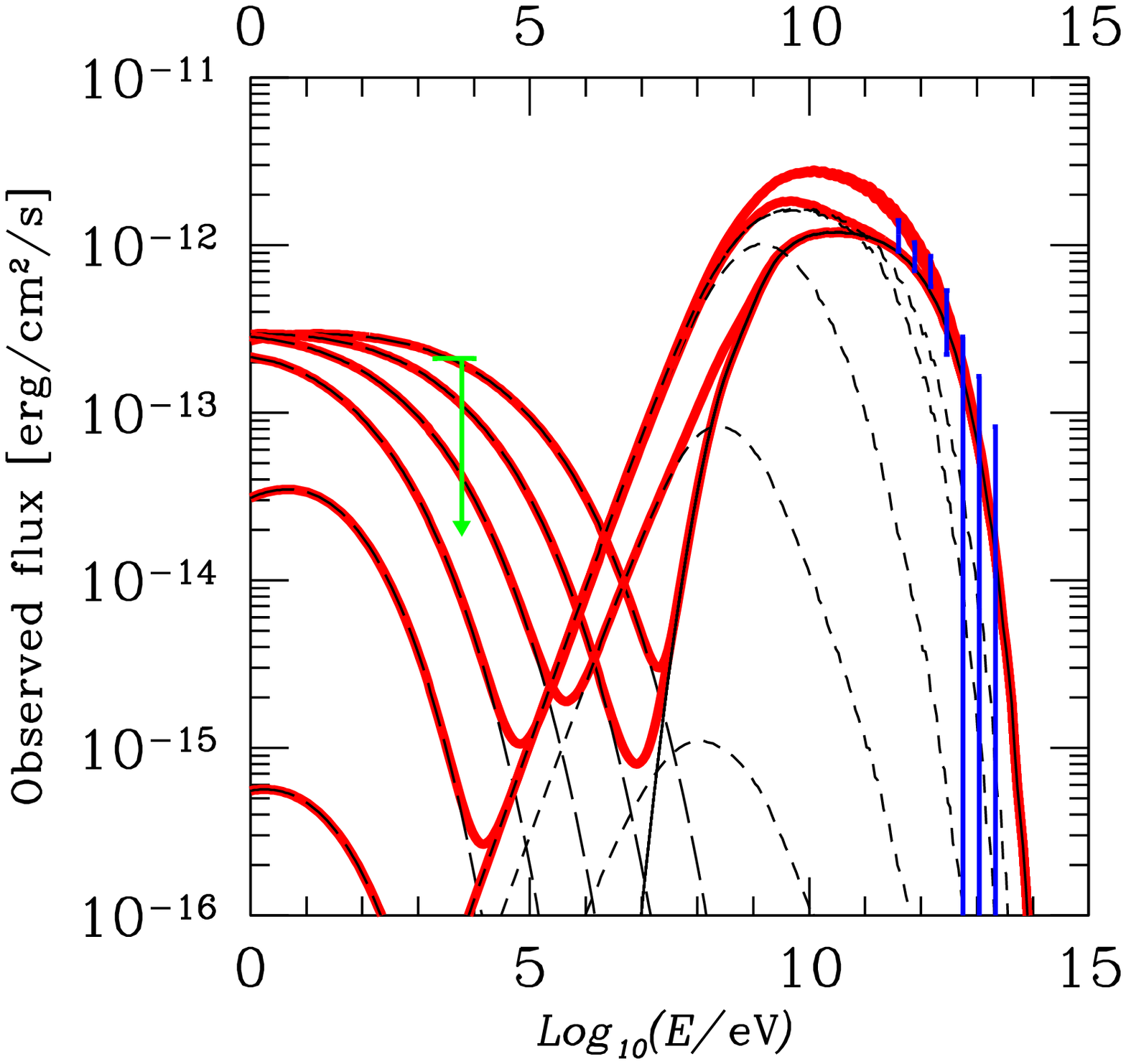}
\caption{
The wide-band $\nu F_\nu$ spectra of the region A of \hess
calculated by the hadronic gamma-ray emission model.
The thick-solid line shows the total nonthermal flux.
Thin-solid line corresponds to $\pi^0$-decay gamma-rays.
The  dotted, dashed, and long-dashed lines are
for the inverse-Compton, bremsstrahlung, and
synchrotron emissions
from secondary electrons and positrons produced by charged pions.
The left panel is for the low-target density model
($n=1$~cm$^{-3}$), while the right is for the high-target
density model ($n=5\times10^3$~cm$^{-3}$).
In calculating the synchrotron radiation of secondary
electrons and positrons,
the magnetic filed strength is changed;
1~G, 100~mG, 10~mG, 1~mG, 100~$\mu$G,
and 10~$\mu$G downward from right to left.
In both panels, we assume that the primary proton spectrum
is in the form of $E^{-p}\exp(-E/E_{\rm max,p})$ 
with an index $p=2.0$ and the 
cut-off energy, $E_{\rm max, p}=18$~TeV.
}
\label{fig:spectrum}
\end{figure}

At first we consider the case $n=1$~cm$^{-3}$
(see the left panel of Fig.~\ref{fig:spectrum}).
Since the value of $n$ is small, emissions from 
secondary electrons is dim in the gamma-ray band, so that
the $\pi^0$-decay  emission dominates.
We find that as long as $n\lesssim10^2$~cm$^{-3}$,
the bremsstrahlung emission of secondary electrons and 
positrons is not dominant in the gamma-ray band.
We fix $E_{\rm max, p}=18$~TeV in order to fit the observed
 spectral shape in the TeV energy range.
Then, as can be seen in the left panel of 
Fig.~\ref{fig:spectrum},
derived upper limit on nonthermal X-rays are explained
if the magnetic field $B\lesssim100$~mG.
This is rather loose limit on the magnetic field.
Because of the small value of $E_{\rm max, p}$,
the maximum energy of secondary electrons and positrons
is small, $\sim1$~TeV, so that
the large value of $B$ is necessary for bright
synchrotron X-rays emitted by them.
This comes from the fact that
the characteristic frequency
of electron/positron synchrotron radiation is given by
\[
 h\nu_{\rm syn}\sim6~{\rm keV}
\left(\frac{B}{50~{\rm mG}}\right)
\left(\frac{E_{{\rm e}^\pm}}{1~{\rm TeV}}\right)^2~~,
\]
where $E_{{\rm e}^\pm}$ is the energy of secondary
electrons or positrons.
For assumed distance of 8.5~kpc,
the total energy of accelerated protons is estimated as
$\sim6\times10^{50}$~erg.

Next we consider the case $n=5\times10^3$~cm$^{-3}$
(see the right panel of Fig.~\ref{fig:spectrum}).
If the magnetic field strength is small 
($B\lesssim10^2~\mu$G),
the bremsstrahlung emission from secondary 
electrons and positrons is
comparable to the $\pi^0$-decay gamma-ray emission.
However, in the strong field cases,
the synchrotron cooling effect significantly suppresses
the number of $e^\pm$, resulting small contribution
to the gamma-rays.
On the other hand, the synchrotron radiation of
secondary $e^\pm$ is brighter for larger $B$.
Similar to the case of $n=1$~cm$^{-3}$,
the magnetic field is no larger than $\sim100$~mG.
\citet{robinson96} derived the relatively high
magnetic field value, 0.2--0.6~mG, which is measured
from the Zeeman splitting of maser lines.
The molecular cloud associated with the region~A
might have the similar magnetic field strength,
and then, $\sim$mG field and the number density
of the cloud $n\sim5\times10^3$~cm$^{-3}$
is roughly consistent
with the previous result on other  molecular
clouds \citep{crutcher91}.
As long as the field strength in the cloud is
$B\sim$~mG or less, the expected flux of secondary synchrotron 
radiation in the X-ray band is rather small and below
the observed upper limit.
For assumed distance of 8.5~kpc,
the total energy of accelerated protons is estimated as
$\sim1.2\times10^{47}$~erg.
Then, the mean energy density of high-energy protons
at region~A of \hess is calculated as $\sim6.0$~eV~cm$^{-3}$
if we assume the volume of the molecular cloud of
$1.2\times10^{58}$~cm$^3$ \citep{aharonian2008},

\subsubsection{Leptonic gamma-ray emission model}

In the leptonic model, TeV gamma-rays 
arise via bremsstrahlung and inverse Compton process
(with CMB seed photons)
emissions of  primarily accelerated electrons, and
X-rays arise from their synchrotron radiation.
Again we assume that the electron energy spectrum
has a form of 
$E^{-p}\exp(-E/E_{\rm max,e})$ with an fixed index, $p=2.0$, and the 
cut-off energy, $E_{\rm max, e}$.
As well as the hadronic model, both low- and high-$n$ cases
are discussed.

\begin{figure}
\epsscale{0.45}
\plotone{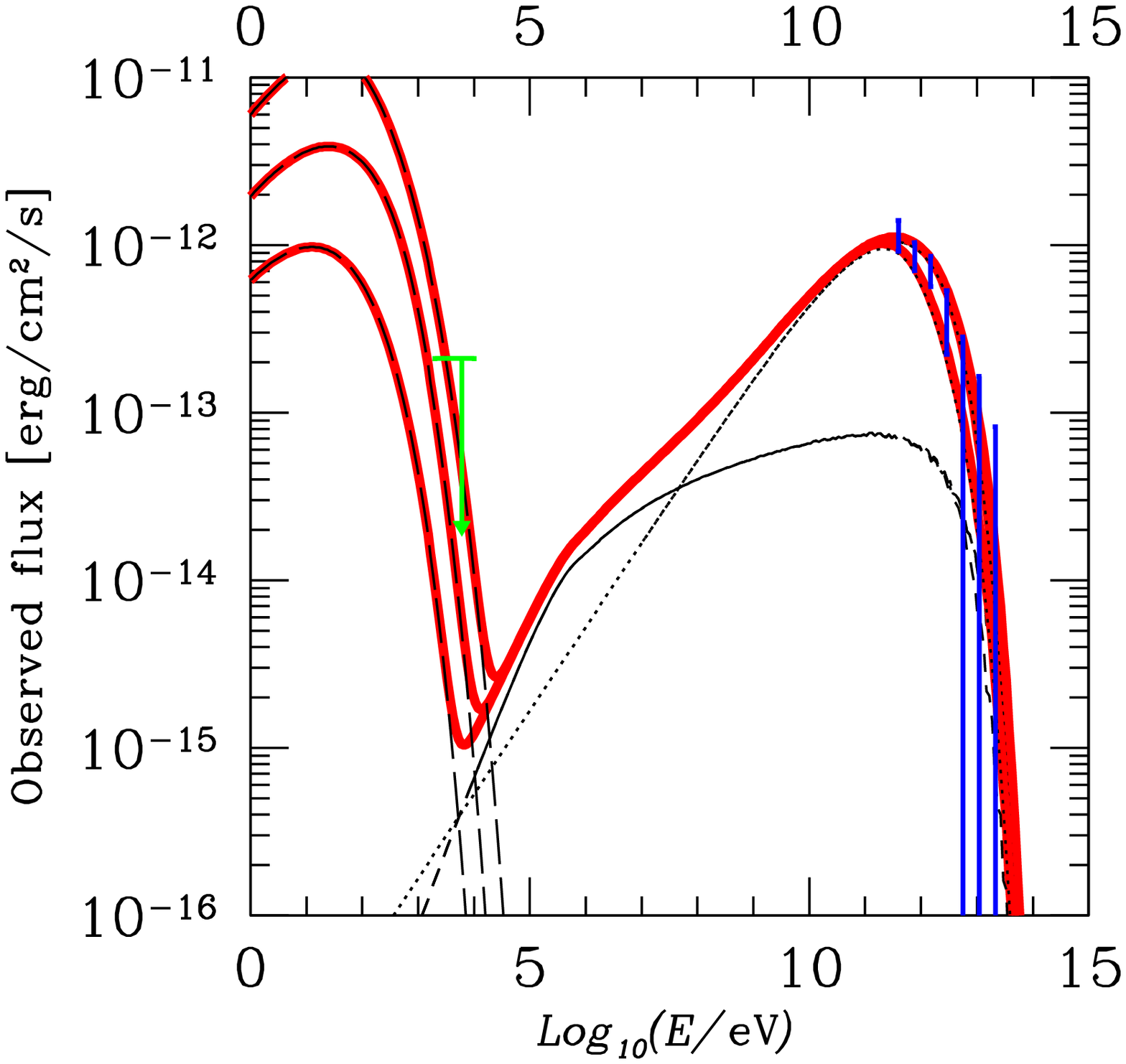}
\plotone{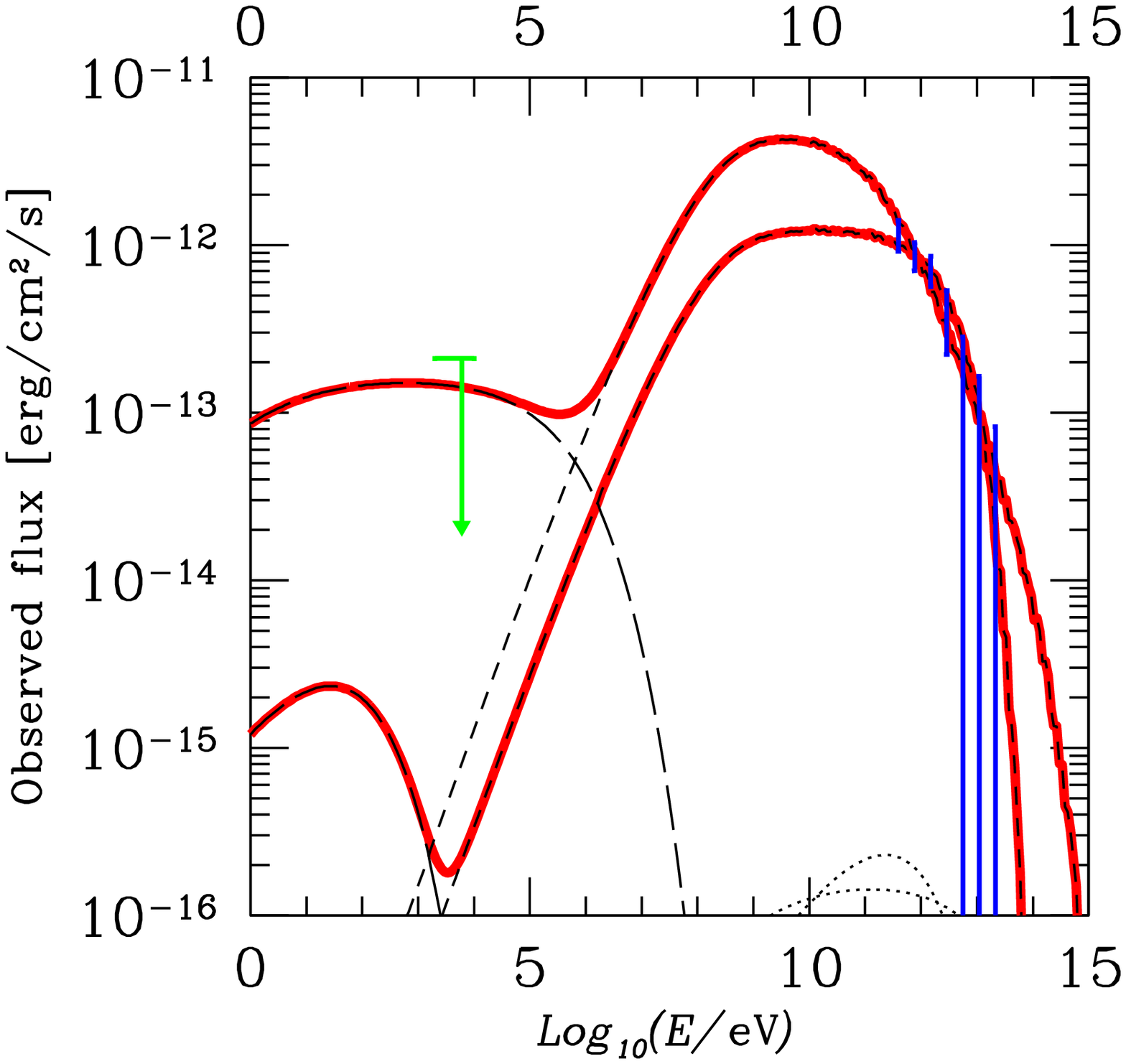}
\caption{
The wide-band $\nu F_\nu$ spectra of the region A of \hess
calculated by the leptonic gamma-ray emission model.
The thick-solid line shows the total nonthermal flux.
The  dotted, dashed, and long-dashed lines are
for the inverse-Compton, bremsstrahlung, and
synchrotron emissions from primarily accelerated electrons.
The left panel is for the low-target density model
($n=0.1$~cm$^{-3}$).
In calculating the synchrotron radiation of secondary
electrons and positrons,
the magnetic filed strength is changed;
12~$\mu$G,
6~$\mu$G, and 3~$\mu$G, downward from right to left.
The right panel is for the high-target density model
($n=5\times10^3$~cm$^{-3}$).
We consider two cases of $B=10$~$\mu$G and $100$~$\mu$G.
In both panels, we assume that the primary electron spectrum
is in the form of $E^{-p}\exp(-E/E_{\rm max,e})$ 
with an index $p=2.0$ and the 
cut-off energy, $E_{\rm max, e}=10$~TeV.
}
\label{fig:spectrum2}
\end{figure}

First we consider the case $n=0.1$~cm$^{-3}$
(see the left panel of Fig.~\ref{fig:spectrum2}).
In this case, the TeV gamma-ray band is dominated by the
IC emission.
In order to explain the observed hard TeV spectrum,
we adopt  $E_{\rm max, e}=10$~TeV.
Then, the relatively weak magnetic field, $B\lesssim6$~$\mu$G,
 is necessary in order for the flux  of synchrotron X-rays
to be smaller than the observed upper limit.
We find that for an assumed distance of 8.5~kpc,
the total energy of accelerated electrons is estimated as
$\sim6\times10^{48}$~erg, which is very large compared with
the typical young SNRs \citep[e.g.,][]{bamba2003b}.

Next we consider the case $n=5\times10^3$~cm$^{-3}$
(see the right panel of Fig.~\ref{fig:spectrum2}).
In this case, the TeV gamma-ray band is dominated by the
bremsstrahlung emission.
When the magnetic field is larger than $\sim100~\mu$G,
the synchrotron cooling effect is significant and
the electron spectrum is modified enough for the emission
spectrum in the TeV gamma-ray band to deviate
from the observed one.
As long as $B\lesssim10^2$~$\mu$G,
we can fit the TeV gamma-ray spectrum with
bremsstrahlung emission, and the synchrotron X-rays
below the observed upper limit.
For assumed distance of 8.5~kpc,
the total energy of accelerated electrons is estimated as
$\sim6.0\times10^{45}$~erg and $\sim2.3\times10^{46}$~erg 
for $B=10$~$\mu$G and $B=100$~$\mu$G, respectively.
If we assume the volume of the molecular cloud of
$1.2\times10^{58}$~cm$^3$ \citep{aharonian2008},
the mean energy density of high-energy protons
at region~A of \hess is calculated as $\sim0.31$~eV~cm$^{-3}$
and $\sim1.2$~eV~cm$^{-3}$ for $B=10$~$\mu$G and $B=100$~$\mu$G, 
respectively.


In the case of $n=5\times10^3$~cm$^{-3}$,
the energy-loss time scale, $t_{\rm loss}$, 
for TeV-emitting electrons via
synchrotron and bremsstrahlung radiation is 
less than $\sim3\times10^3$~yr.
Then, the diffusion length scale is estimated as 
$\sqrt{2Kt_{\rm loss}}\lesssim0.26\eta^{1/2}
(B/10~\mu{\rm G})^{-1/2}
(E_e/{\rm TeV})^{1/2}$~pc,
where we assume the Bohm type diffusion,
$K=cE_e\eta/3eB$.
Even if $\eta\sim10^2$,
this length scale is much smaller than the
source size of region~A of \hess
($\sim40$~pc assuming the distance of 8.5~kpc),
which implies that the high-energy electrons 
coming from a nearby accelerator
cannot penetrate into deep interior of a molecular cloud.
On the other hand, when $n=0.1$~cm$^{-3}$,
relatively large energy of accelerated electrons is required.
So, the leptonic models considered above may be unlikely.
However, in order to derive the  firm conclusion, we need
to discuss further detail with 
more information on \hess, such as
the intrinsic source size, the number density of
target matter, more precise spectrum in all wavelength 
(from radio to GeV--TeV gamma-rays), and so on.

\subsection{Source of High Energy Particles}

Spatial coincidence of a molecular cloud and the TeV gamma-rays
at region~A of HESS~J1745$-$303
suggests that the TeV emission is possibly due to accelearated protons
interacting the molecular cloud.
The source of high energy protons is still unknown.
The past GC activity might be able to make protons
\citep{aharonian2006b}.
However, it is rather difficult due to the low mass of the molecular cloud
in our region.
Nearby SNRs and pulsars, on the other hand, could be high-energy
proton injectors.
G359.1$-$0.5 is an energetic SNR and interacting with the molecular cloud
\citep{uchida1992} and is located in the GC region.
We confirmed that
the distances to this SNR and the molecular clouds are similar,
but we could not find any signature of the interaction
in X-rays.
Hidden SNRs by the molecular cloud
also could be the source of accelerated particles.
The average absorption column of the molecular cloud is
$\sim 10^{23}$~cm$^{-2}$,
which is enough to hide the soft X-ray emission from the SNRs.
For further study, more detailed radio observations are needed.
G359.0$-$0.9 is an unrelated source
with its significantly smaller distance \citep{bamba2000}.
A pulsar B1742$-$30 has been argued as a possible cosmic ray injector
\citep[e.g.,][]{lemiere2008},
but the estimated distance \citep[2.1~kpc;][]{taylor93}
is significantly smaller than that of \hess.
We thus concluded that the pulsar B1742$-$30
is also unrelated to \hess.

\section{Summary}

Deep mapping observations of a part of \hess 
have been carried out with {\it Suzaku}.
The main part of \hess, region~A, is entirely covered
in the present observation.
We made a tight upper limit of hard X-rays of the \hess region,
$2.1\times 10^{-13}$~ergs~s$^{-1}$cm$^{-2}$
in the 2--10~keV band.
The flux ratio between 1--10~TeV and 2--10~keV is larger than 4,
which could make \hess one of the ``dark particle accelerators''.
An excess of the neutral iron emission line on this region
is found.
This possible excess should be due to the X-ray reflection
from a molecular cloud by the past active GC.
This provides strong evidence for the interaction between 
 the molecular cloud --- \hess system in the GC region.
With the small flux in the hard X-ray band and coincidence with
a molecular cloud,
the TeV emission from \hess could be due to protons encountered 
the molecular cloud.

\acknowledgments
We would like to thank the referee for useful comments and
suggestions.
The authors also thank H.~Murakami,
A.~Kawachi, T.~Oka, and Y.~Fukui, for their useful comments
on molecular clouds and X-ray reflection nebulae.
This work was supported in part by
 Grant-in-Aid for Scientific Research
of the Japanese Ministry of Education, Culture, Sports, Science
and Technology, No.~19$\cdot$4014 (A.~Bamba), 
No.~18740153 and No.~19047004 (R.~Yamazaki),
and also supported in part by PPARC grant, PP/D000394/1, EU grant
MRTN-CT-2006-035863, the European Union through the Marie Curie
Research and Training Network ``UniverseNet'', 
MRTN-CT-2006-035863 (K.~Kohri).



\begin{deluxetable}{p{5pc}cccc}
\tabletypesize{\scriptsize}
\tablecaption{Observation Log
\label{tab:obslog}}
\tablewidth{0pt}
\tablehead{
\colhead{ObsID} & \colhead{Date} & \colhead{Position} & \colhead{Exposure} & \colhead{SCI} \\
 & YYYY-MM-DD & (J2000) & [ksec]}
\startdata
501010010\dotfill & 2006-10-07 & (266.26, $-$30.37) & 53 & OFF \\
502016010\dotfill & 2008-03-02 & (266.22, $-$30.11) & 73 & ON \\
502017010\dotfill & 2008-03-06 & (266.47, $-$30.09) & 75 & ON \\
502018010\dotfill & 2008-03-08 & (266.06, $-$30.24) & 82 & ON
\enddata
\end{deluxetable}

\begin{deluxetable}{p{15pc}ccccc}
\tabletypesize{\scriptsize}
\tablecaption{Comparison of The Iron Emission Lines
\label{tab:iron}}
\tablewidth{0pt}
\tablehead{
& \colhead{Src} & \colhead{Bgd1} & \colhead{Bgd2} & \colhead{Bgd3} & \colhead{Bgd average}
}
\startdata
Neutral Iron\\
\hspace*{2mm}
Line center [eV]\dotfill & 6.42 (6.41--6.43) & (fixed to Src1) & 6.38 (6.36--6.41) & 6.41 (6.37--6.45) & 6.40 (6.38--6.42) \\
\hspace*{2mm}
Surface brightness [$10^{-8}$ph~cm$^{-2}$s$^{-1}$arcmin$^{-2}$]\dotfill & 6.5 (5.8--7.2) & 5.9 (3.7-8.1) & 6.6 (4.3--8.9) & 2.3 (1.2--3.4) & 3.2 (2.5--3.9) \\
He-like Iron\\
\hspace*{2mm}
Line center [eV]\dotfill & 6.68 (6.67--6.69) & (fixed to Src1) & 6.67 (6.66--6.68) & 6.69 (6.67--6.71) & 6.68 (6.67--6.69) \\
\hspace*{2mm}
Surface brightness [$10^{-8}$ph~cm$^{-2}$s$^{-1}$arcmin$^{-2}$]\dotfill & 11 (10--12) & 12 (10--14) & 17 (14--20) & 10 (9--11) & 11 (10--12)
\enddata
\end{deluxetable}

\begin{deluxetable}{p{10pc}cc}
\tabletypesize{\scriptsize}
\tablecaption{Best-fit parameters of G359.0$-$0.9 and G359.1$-$0.5%
\tablenotemark{a}.
\label{tab:snrs}}
\tablewidth{0pt}
\tablehead{
& \colhead{G359.0$-$0.9} & \colhead{G359.1$-$0.5}
}
\startdata
$kT_1$ [keV]\dotfill & 0.35 (0.32--0.38) & 1.1 (0.7--2.0) \\
Mg \dotfill & 1.4 (1.1--1.6) & ---\tablenotemark{b} \\
Si \dotfill & ---\tablenotemark{b} & 1.9 (1.3--2.9) \\
Fe \dotfill & 0.4 (0.2--0.6) & ---\tablenotemark{b} \\
$n_et$ [$10^{11}$~cm$^{-3}$s]\dotfill & 2.3 (1.5--3.7) & ---\tablenotemark{c} \\
$kT_2$ [keV]\dotfill & ---\tablenotemark{d} & 2.0 (1.7--2.5) \\
S \dotfill & ---\tablenotemark{d} & ($>$40) \\
$N_{\rm H}$ [$10^{22}$~cm$^{-2}$]\dotfill & 1.6 (1.5--1.8) & 2.3 (1.6--2.5) \\
Flux (2--10~keV) [ergs~s$^{-1}$cm$^{-2}$]\dotfill & $7.1\times 10^{-14}$ & $2.6\times 10^{-13}$ \\
$\chi^2$/d.o.f.\dotfill & 109.7/70 & 64.0/41
\enddata
\tablenotetext{a}{Errors indicate single parameter 90\% confidence regions.}
\tablenotetext{b}{Fixed to the solar abundance.}
\tablenotetext{c}{Not determined.}
\tablenotetext{d}{This component was not used for this SNR.}
\end{deluxetable}

\begin{deluxetable}{p{9pc}cc}
\tabletypesize{\scriptsize}
\tablecaption{Characteristics of molecular clouds
\label{tab:MC}}
\tablewidth{0pt}
\tablehead{
\colhead{} & \colhead{Sgr~B2\tablenotemark{a}} &
\colhead{HESS~J1745$-$303\tablenotemark{b}}
}
\startdata
$M_{\rm MC}$\tablenotemark{c}~[$M_\odot$]\dotfill & $6\times 10^6$ & $5\times 10^4$ \\
$\theta$~[deg.]\tablenotemark{d}\dotfill & 0.05 & 0.3 \\
$D$\tablenotemark{e}~[deg.]\dotfill & 1.2 & 0.7
\enddata
\tablenotetext{a}{\cite{murakami2001}.}
\tablenotetext{b}{\cite{aharonian2008}.}
\tablenotetext{c}{Mass of the molecular cloud.}
\tablenotetext{d}{Angular size of the cloud from the direction of the GC.}
\tablenotetext{e}{Angular separation from the GC to the cloud.}
\end{deluxetable}

\end{document}